\centerline{Derivation of the Pauli exchange principle}
\centerline{A. A. Broyles}
\centerline{Department of Physics, Institute for Fundamental
Theory}
\centerline{University of Florida, Gainesville, Fl 32611}

When the time-independent wave function of a multicomponent system
is computed, it is necessary to require that it be consistent with
the Pauli exchange principle which states that, for systems
involving identical elements, wave functions must be
{\it antisymmetric} to pair exchange if twice the spin $s$ for each
one is an odd integer and {\it symmetric} to pair exchange if the
spin is an integer. For identical spin sets, each set containing
spin $s$, this can be expressed mathematically, for a single
exchange, in the form,  
$$\Psi({\bf x}_1{\underline s} \cdots {\bf x}_a{\underline s}
\cdots {\bf x}_b {\underline s}\cdots {\bf x}_n{\underline s})=
(-1)^{2s}\Psi({\bf x}_1{\underline s} \cdots {\bf x}_b{\underline
s}\cdots {\bf x}_a {\underline s} \cdots{\bf x}_n{\underline s}) 
\eqno(1)$$
where ${\bf x}$ is a set of space coordinates, and ${\underline s}$
represents a set of spin quantum numbers.

Pauli [1] showed that this relation connecting fermion and boson
statistics with spin can be derived with the aid of relativistic
quantum field theory. As will be seen below, it can also be derived
directly from the properties of wave functions. Papers have been
published by Feynman, Schwinger, and others in an effort to find a
simpler and more understandable proof. Their work is discussed in
a recent review article [2] in which we find this statement,
``Finally we are forced to conclude that although the
Spin-Statistics Theorem is simply stated, it is by no means simply
understood or simply proved.''

Feynman's position on the lack of a simple derivation in 1963 was
expressed in his physics course [3] where he said, ``This probably
means we do not have a complete understanding of the fundamental 
principle involved.''  Broyles [4] published an elementary proof of
this theorem in 1976 in which he used an exchange operator defined
in terms of rotation operators. Apparently unaware of this, Feynman
presented a lecture [5] giving essentially the same proof in 1986.
However, Duck and Sudarshan [2] objected to these proofs by
expressing the need for evidence that the wave functions before and
after the exchange are identical. This paper is written to answer
this objection and to summarize the proof.

Feynman [6] introduced the term, ``probability amplitude'',
into quantum mechanics. The square of the magnitude of this complex
number gives the probability of obtaining a specified collection of
quantum numbers as the result of measurements. In order to identify
this amplitude uniquely (aside from an overall constant phase), a
{\it complete} set of quantum numbers must be specified. As an
example, Kemble [7] relates how he and Feenberg proved that the
probability amplitude over all space for position coordinates can
be determined by repeated measurements. These measurements can be
used to construct the probability distribution and its time
derivative. They show that a phase factor, constant in space and
time, remains undetermined. Phases of probability amplitudes can be
measured relative to a standard by means of diffraction experiments
[8], but no method is available for measuring the phase of the
standard. We shall assume here that a complete set of quantum
numbers can be found to determine {\it any} probability amplitude
to within a constant phase.

These quantum numbers are associated in sets as though each set
belonged to a particle. Each set can include the configuration
space coordinates ${\bf x}$ as well as the spin coordinates in the
subset ${\underline s}$. We shall include the spin $s$ and a spin
component $s_z$ in ${\underline s}$. In addition, any other quantum
numbers such as mass, charge, flavor, etc. that may be needed to
identify the set, can be included in ${\underline s}$. In
association with each set ${\bf x}{\underline s}$ , we can imagine
a point in a representative spin©configuration space whose
coordinates are the components of ${\bf x}{\underline s}$. In
Fig.~1a, points representing three sets, that we shall take to have
identical spin quantum numbers ${\underline s}$ , are pictured. We
shall be interested in systems where all the sets ${\underline s}$
are identical except possibly for the $s_z$ quantum numbers. Let us
first consider cases where all the $s_z$'s are the same and where
each one has its maximum value $s$.

The wave function $\Psi$ in Eq.~(1) is a convenient way to write
down this probability amplitude and its associated sets of quantum
numbers. However it is clear that additional information has been
incorporated in this wave function, additional information beyond
that contained in the quantum numbers that determine the probably
amplitude. This information is provided by the {\it arrangement} of
the sets in the argument of the wave function. The number
represented by the function $\Psi$ will, in general, depend upon
the {\it order} in which the sets of quantum numbers are written.
The probability amplitude, on the other hand, depends only on what
quantum number sets are present, not on the order in which they
appear. Thus if the wave function is to represent the probability
amplitude, we must impose a restriction on it to remove the
dependence on the order. If the spins are zero, all we need to do
is make the wave function symmetric to the pair exchange of
identical sets. For nonvanishing spins, however, the restriction is
more complicated.

It is instructive to recognize a difference between the spatial
exchange of two apparently identical {\it classical} objects and
the exchange of two {\it electron} coordinate sets determining a
probability amplitude. For example if we exchange two billiard
balls, we produce a new state, a new configuration. Of course, if
the two balls had exactly the same number of atoms of each kind, it
might be possible to make them truly indistinguishable so that the
state {\it after} the exchange is the same as the one before.
Since, however, it is essentially impossible to equalize the number
of atoms in the two balls, the probabilities are overwhelming that
the initial and final states will be different. If however, two
{\it electron} coordinate sets, ${\bf x}_a {\underline s}$ and
${\bf x}_b{\underline s}$, with identical ${\underline s}$ subsets,
are exchanged, the quantum numbers present are the same after the
exchange as before. {\it Thus the probability amplitude, since it
depends only on the sets that are present and not on their order,
will be unaltered by the exchange.} This is illustrated in Fig.~1
where the exchange in Figs~1b and 1c leaves Fig~1d identical to 1a.
This is equivalent to saying that the ${\bf x}{\underline s}$'s
that specify a probability amplitude are {\it generic} sets rather
than {\it specific} [9].

The argument of a wave function, is normally written down with the
quantum number sets in specific positions. Thus there is one set in
the first position to which we can assign a number one and label
the symbols with subscripts so that we have ${\bf x}_1{\underline
s}_1$. Similarly, the particular set in the second position can be
written as ${\bf x}_2{\underline s}_2$, etc. Then an application of
the exchange operator to sets $a$ and $b$
changes their positions in the argument of the wave function as we
see in Eq.~(1). In order to make the wave function equal the
probability amplitude, it must not be allowed to change value as a
result of this exchange.

To generate the N! {\it specific} positions that can appear in the
argument of the wave function from the values of the wave function
at the N {\it generic} positions that are available from the
probability amplitudes, we must first identify each {\it generic}
set. This can be done by assigning an order number to it. One way
of accomplishing this is to select a point in the representation
space to act as the center. The generic positions can then be
numbered according to their distances from this center. The point
nearest to the center will be point number one, the next nearest
will be point number two, etc. The distance from the center to the
nearest point can then be represented by $r_1$, to the next nearest
one by $r_2$, etc. Then we have $r_1<r_2<r_3<\cdots r_n$. The
probability amplitudes for these N points then compose the wave
function for the region where the $r$'s satisfy these inequalities.
 We shall call this, Region A.

Since the above procedure gives $\Psi$ in a sizeable fraction of
space, Region A, the assumption is made here that the $\Psi$ in the
remainder of space (where the $r$'s satisfy different inequalities)
can be computed by {\it analytic continuation}. Since any order of
$r$'s can be altered to a {\it new} order by exchanging two points
in the representation space, it is convenient to obtain the
collection of Taylor's series for this continuation by expanding
the rotation operators that move two points through angles. Such an
exchange is illustrated in Figs.~1b and 1c. If it exchanges the
points labeled 1 and 2 in the last paragraph, the distances from
the center in the representation space are in the order
$r_2<r_1<r_3\cdots r_n$. These points are in a region other than
Region A.

If the sets to be exchanged are ${\bf x}_a{\underline s}_a$ and
${\bf x}_b{\underline s}_b$, the rotations required can be
described by first placing the representation space coordinate
system so that its $x$ axis coincides with the line connecting the
two points at ${\bf x}_a$ and ${\bf x}_b$ and with the origin
(marked O in Figs.~1b and 1c) half way between these points. Then
the operator [10] (with the subscript $a,b$ indicating ``$a$ or
$b$''),
$$R^{\phi}_{a,b}=e^{i\phi_{a,b} J_{za,b}} \eqno(2)$$
where 
$$J_{za,b}=L_{za,b}+S_{za,b} \eqno(3)$$
and
$$L_{za,b}=i\hbar {\partial\over \partial \phi_{a,b}}\ ,\eqno(4)$$
($S_{za,b}$ is the generator of the spin component in the $z$
direction.) can be applied to $\Psi$ to produce the $\Psi$ after
the required rotations. This rotation operator or its series
can be used to compute the wave function with ${\bf x}_a$ or
${\bf x}_b$ moved by a rotation through an angle $\phi_{a,b}$. If
the spins in the argument of $\Psi$ are zero, the spin generator in
Eq.~(3) can be omitted. However, if the spins do not vanish, the
spin generator will be required since the direction of spin in the
argument of $\Psi$ after the rotation may be different from the
initial one. 

When $\phi_a$ equals $\pi$, $\Psi$ will be calculated at the points
resulting from the rotation of ${\bf x}_a{\underline s}_a$ as shown
in Fig.~1b. Next, setting $\phi_b$ to $\pi$ provides the $\Psi$
after the rotation of ${\bf x}_b,{\underline s}_b$ through $\pi$ as
shown in Fig~1c. The two rotations together $R_aR_b$ then produce
the $\Psi$ with the two points {\it exchanged}. See References [4]
and [5] for additional discussions. By applying the proper choice
of these rotation operators to $\Psi$, the wave function with
{\it any} exchange of sets can be computed. By computing the proper
choice of exchanges, the wave function for any specific
configuration of sets can be determined from a generic one.

In the case under consideration where the spin components are
identical, that is, $s_{za}=s_{zb}=s$, the $L_z$ operators will
exchange the ${\bf x}$'s while the $S_z$'s can be replaced by their
eigenvalues. As a result, the two spin rotation operators applied
to $\Psi$ to produce an exchange will give a factor,
$e^{i2s_{za}\pi}=(-1)^{2s}$.
Thus
$$R_aR_b \Psi(\cdots {\bf x}_a{\underline s}\cdots {\bf x}_b
{\underline s}\cdots)=(-1)^{2s}\Psi(\cdots {\bf x}_b{\underline s}
\cdots {\bf x}_a{\underline s}\cdots) \ . \eqno(5)$$
where the $\phi_{a,b}$ in Eq.~(2) equals $\pi$ in $R_{a,b}$.

But, as we have seen, exchanging these coordinate sets does not
introduce new sets or remove old ones and, therefore, does not
change the probability amplitude. Thus $R_aR_b$ must have no effect
on $\Psi$ if $\Psi$ is to equal the probability amplitude. As a
result, the left hand side is equal to $\Psi(\cdots
{\bf x}_a{\underline s}\cdots {\bf x}_b {\underline s}\cdots)$.
This makes the last equation equivalent to Eq.~(1). The effect of
exchanging two points lying on a node can be computed by a limiting
procedure.

If the quantum numbers in the various spin sets in the argument of
$\Psi$ are identical {\it except} for the values of $s_z$, the
Pauli principle again applies. To prove this, it is convenient to
make use of the complete set of spin wave functions (defined with
the aid of an integer $l$) with the property,
 $$\chi(s,\theta_l)=e^{i\theta_l S_x}\chi(s,0),\ \ \   
\theta_l=l\pi/2s,\ \ \  0\le l\le 2s, \eqno(6)$$
introduced in the appendix of Reference [4]. The $\theta_l$'s then
replace the $s_z$'s in the argument of the wave function. A wave
function like that in Eq.(1) can then be written in the form,
$$\Psi(\cdots{\bf x}_as\theta_{la}\cdots {\bf x}_bs\theta_{lb} 
\cdots) =\cdots e^{i\theta_{la} S_{xa}} \cdots e^{i\theta_{lb}
S_{xb}} \cdots \Psi(\cdots {\bf x}_as0 \cdots {\bf x}_bs0 \cdots)
\eqno(7)$$
where the zero's indicate values of the $\theta_l$'s. It is
understood that the subscript $j$ on $S_{xj}$ is the same as the
subscript on the ${\bf x}_j$ in the set on which it operates.
Since, as we have seen in the last section, $\Psi(\cdots
{\bf x}_as0 \cdots {\bf x}_bs0 \cdots)$ satisfies Eq.(1), and the
operators, $e^{i\theta_{la} S_{xa}},e^{i\theta_{lb}S_{xb}},\cdots$
commute, it follows that $\Psi( \cdots {\bf x}_as \theta_a \cdots
{\bf x}_bs\theta_b \cdots)$ will also satisfy Eq.(1). 

The author is indebted to J. R. Klauder, R. L. Coldwell, C. B.
Thorn, and H. P. Hanson for discussions about this manuscript.
\vfill\eject

\centerline{\bf REFERENCES}
\vskip .25in

\item{[1]} W. Pauli, Phys. Rev. {\bf 58}, 716
(1940). 

\item{[2]} I. Duck and E. C. G. Sudarshan, Am. J. Phys.
{\bf 66}, 284 (1998).

\item{[3]} R. P. Feynman, R. B. Leighton, and M. Sands, 
{\it The Feynman Lectures on Physics} (Addison-Wesley,
Reading, Mass., 1965) Vol. 3, p. 162. 

\item{[4]} A. A. Broyles, Am. J. Phys. {\bf 44}, 340
(1976).

\item{[5]}, ``The Reason for Antiparticles'' in R. P. Feynman
and S. Weinberg, {\it Elementary Particles and
the Laws of Physics} (Cambridge University Press, New York,
1987).

\item{[6]} R. P. Feynman, {\it The Theory of Fundamental
Processes} (W. A. Benjamin, New York,  1962),
Chapter 1. 

\item{[7]} E. C. Kemble, {\it The Fundamental Principles of
Quantum Mechanics} (McGraw-Hill, New York,
1937), First Edition, p. 71.

\item{[8]} J. J. Sakurai, {\it Modern Quantum Mechanics}
(Benjamin/Cummings, Menlo Park, CA) 
p. 162.

\item{[9]} D. ter Haar, {\it Elements of Statistical Mechanics}
(Rinehart, New York, 1954), p. 136,  last
paragraph.

\item{[10]} Ramamurti Shankar, {\it Principles of Quantum
Mechanics} (Plenum Press, New York, 1980)
Chapter Sec. 12.5.\vfill\eject
 
\vfill

\eject
\null \ 
\input psfig
\psfig{figure=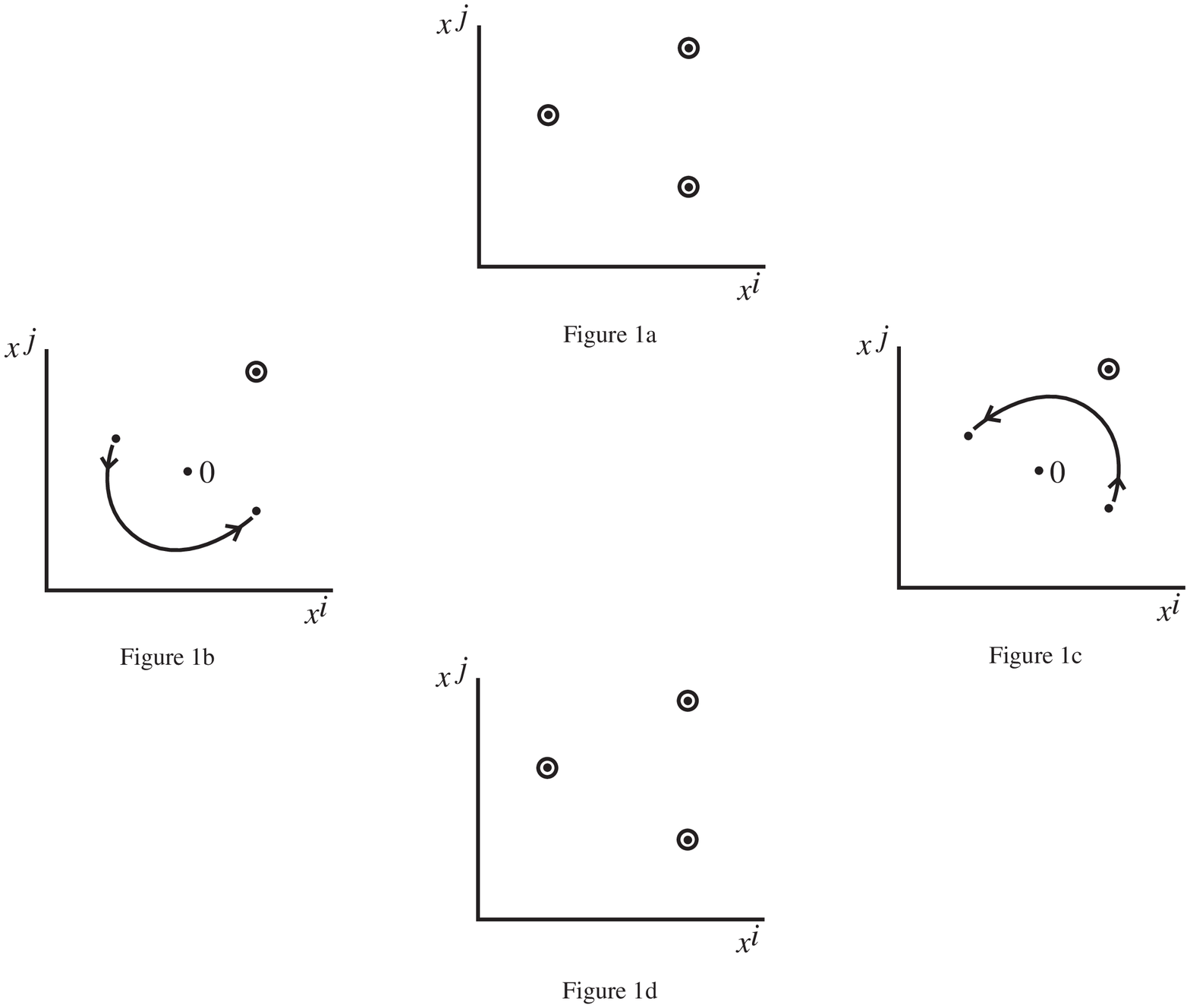,width=6in}

\vskip .5in

\item{Figure 1.} The exchange, by means of rotations, of two points
representing two sets of quantum numbers. Labels are unnecessary
because we take the spin quantum numbers to be identical. The
coordinate axes are located so that the origin is half way between
the two points to be exchanged. 1a. The projection into the $x©y$
plane of three points representing three sets of quantum numbers.
1b. The rotation of one of the points through an angle $\pi$ around
the $z$ axis with the origin at $O$. 1c. The rotation of the other
point through an angle $\pi$. 1d. The representative points after
the two rotations that result in the exchange. Fig.~1d is identical
to 1a.

\end